\documentclass[conference]{IEEEtran}
\IEEEoverridecommandlockouts

\usepackage[T1]{fontenc}
\usepackage[utf8]{inputenc}
\usepackage{cite}
\usepackage{amsmath,amssymb,amsfonts}
\usepackage{graphicx}
\usepackage{textcomp}
\usepackage{xcolor}
\usepackage{booktabs}
\usepackage{array}
\usepackage{url}
\usepackage{multirow}
\usepackage[nolist]{acronym}
\usepackage{subcaption}
\usepackage{footmisc}
\usepackage[font={small,it},labelfont={small,bf,it}]{caption}
\begin{document}

\begin{acronym}[CAEX]
\acro{fpd}[FPD]{Formalized Process Description}
\acro{aml}[AML]{AutomationML}
\acro{caex}[CAEX]{Computer Aided Engineering Exchange}
\acro{dd}[DD]{Diagram Definition}
\end{acronym}

\title{An AutomationML Domain Library for the Formalized Process Description}
\author{
\IEEEauthorblockN{Hamied Nabizada\IEEEauthorrefmark{1}, Rainer Drath\IEEEauthorrefmark{2}, Felix Gehlhoff\IEEEauthorrefmark{1}, Alexander Fay\IEEEauthorrefmark{3}}
\IEEEauthorblockA{\IEEEauthorrefmark{1}Institute of Automation Technology, Helmut Schmidt University Hamburg, Hamburg, Germany}
\IEEEauthorblockA{\IEEEauthorrefmark{2}Institute of Smart Systems and Services, Pforzheim University, Pforzheim, Germany}
\IEEEauthorblockA{\IEEEauthorrefmark{3}Chair of Automation Technology, Ruhr University Bochum, Bochum, Germany}
}

\maketitle

\begin{abstract}
The \ac{fpd} according to VDI/VDE~3682 provides a standardized graphical notation for describing processes across engineering domains but lacks a standardized, tool-independent data format for machine-readable model exchange.
This paper presents an \ac{aml} domain library that formalizes the complete set of \ac{fpd} language elements, their attributes, connection semantics, and graphical representation information as class libraries based on the \ac{caex}~3.0 metamodel.
The library comprises five interrelated parts: a RoleClassLib defining the semantic roles, an InterfaceClassLib for connection types, two AttributeTypeLibs for the information model and diagram interchange, and a SystemUnitClassLib providing instantiation templates.
Key design decisions regarding inheritance, diagram structure, hierarchical decomposition, and the representation of graphical information are discussed along with the alternatives that were considered.
A bidirectional mapping tool demonstrates the library's applicability by converting between a web-based \ac{fpd} modeler and \ac{aml}.

\end{abstract}


\section{Introduction}

The \acf{fpd} according to VDI/VDE~3682~\cite{VDI3682Blatt1} is a standardized graphical notation for describing processes across engineering domains.
It defines six object types and four connection types.
\textit{Product}, \textit{Energy}, and \textit{Information} represent states that serve as inputs and outputs of a \textit{Process Operator}.
A \textit{Technical Resource} realizes one or more process operators via usage relationships, and the \textit{System Limit} separates the process from its environment.
Directed flow connections represent material, energy, and information flows, and the notation supports hierarchical decomposition of process operators into subordinate processes~\cite{VDI3682Blatt1}.
Part~2~\cite{VDI3682Blatt2} complements the graphical notation with an information model that specifies \textit{Identification} and \textit{Characteristics} attributes for each element.

The \ac{fpd} has been adopted in areas such as capability modeling~\cite{Kocher2020}, digital process twins~\cite{Caesar2020}, automated sequence planning~\cite{Vieira2023,Nabizada2026CASE}, and ontology-based process integration in materials engineering~\cite{Schneider2025}.
Despite this growing adoption, the standard does not define a standardized data format for tool-independent model exchange.
Existing tools such as FPB.js\footnote{\url{https://www.fpbjs.net}} and the FPD-Editor\footnote{\url{https://github.com/FLEXCELERATE/fpd-editor}} therefore store process descriptions in tool-specific formats~\cite{Nabizada2020}.
As a result, exchanging \ac{fpd} models between tools requires dedicated mappings or conversions, which limits interoperability and hinders reuse in broader engineering workflows.

\acf{aml}~\cite{IEC62714} is a vendor-neutral data format for exchanging engineering information.
It uses \acf{caex}, originally defined in IEC~62424~\cite{IEC62424}, as its underlying metamodel.
\ac{caex} provides \textit{RoleClasses} for semantic roles, \textit{SystemUnitClasses} as instantiation templates, \textit{InterfaceClasses} for typed connection endpoints, and \textit{Attributes} for properties.
Classes are organized in reusable libraries, while concrete models are represented as InstanceHierarchies~\cite{Drath2021}.
This makes \ac{aml} a suitable foundation for representing graphical description languages by means of reusable domain-specific libraries that can be interpreted by \ac{aml}-compliant tools without modifying the standard itself.

This paper presents an \ac{aml} domain library that formalizes the \ac{fpd} as a set of \ac{caex} class libraries, building on a general methodology for mapping graphical description languages into \ac{aml}~\cite{NabizadaDrath2026}.
The contributions of this paper are: (1)~the complete \ac{aml}/\ac{caex} domain library for the \ac{fpd}, (2)~a discussion of key design decisions and their alternatives, and (3)~a bidirectional mapping tool with round-trip validation.
The library is currently under review in the GMA Technical Committee~2.19 as a candidate for Part~3 of VDI/VDE~3682.

\section{Related Work}
\label{sec:related}

Several approaches to machine-readable representations of the \ac{fpd} have been proposed.
In~\cite{Lueder2010}, \ac{aml} was used to model specific manufacturing processes directly, rather than mapping an existing graphical description language to \ac{aml}. 
In~\cite{Jaeger2012}, an \ac{fpd}-to-\ac{aml} mapping was presented that enabled a toolchain from requirements engineering to plant structure descriptions. While it established an initial conceptual mapping between \ac{fpd} and \ac{caex} elements, the approach was tied to a Microsoft Visio-based workflow and did not produce a reusable, tool-independent domain library.
In~\cite{Nabizada2020}, a web-based modeling tool with a JSON-based exchange format was proposed, and in~\cite{Nabizada2022} an XML representation aligned with the VDI~3682 information model. Both, however, introduce a new domain-specific data format. In~\cite{Beers2024}, a SysML profile was proposed that embeds the \ac{fpd} into an existing modeling framework but targets systems engineering workflows rather than industrial data exchange.
In~\cite{Hildebrandt2018}, the VDI~3682 information model was formalized as an OWL ontology using Semantic Web technologies, leveraging an existing ecosystem but without covering graphical representation information such as element positions and connection routing, so that the original graphical process description cannot be reconstructed from the model.

The approach presented here differs from~\cite{Jaeger2012} by providing a systematic, reusable domain library that is independent of any specific modeling tool.
Unlike~\cite{Nabizada2020, Nabizada2022, Beers2024}, it avoids introducing a new format by building on the established \ac{aml} ecosystem. 
This paper presents the \ac{fpd} in detail and discusses the rationale behind its design decisions.

\section{\ac{fpd} Domain Library}
\label{sec:library}

The domain library consists of five \ac{caex} class libraries: a RoleClassLib for semantic roles, an InterfaceClassLib for connection types, two AttributeTypeLibs for the information model and diagram interchange, and a SystemUnitClassLib for instantiation templates.

\subsection{RoleClassLib -- Semantic Roles}

The \textit{VDI\_FPD\_RoleClassLib} defines the semantic meaning of each \ac{fpd} element.
Fig.~\ref{fig:roleclass} shows the inheritance structure.
\texttt{FPD\_Object} serves as the abstract base class, inheriting from \texttt{AutomationMLBaseRole}.
It carries the shared attributes \texttt{Identification}, \texttt{Characteristics}, and \texttt{ViewInformation} (of type \texttt{FPD\_Bounds}, storing position and dimensions).
Five classes inherit from it: \texttt{FPD\_SystemLimit}, \texttt{FPD\_State} (with sub-roles \texttt{FPD\_Product}, \texttt{FPD\_Energy}, \texttt{FPD\_Information}), \texttt{FPD\_ProcessOperator}~(an individual processing step), and \texttt{FPD\_TechnicalResource}.

\texttt{FPD\_Process} inherits from the \ac{aml} base role \texttt{Structure} rather than from \texttt{FPD\_Object}. 
Unlike the six visible object types, a process serves as the structural container for a diagram rather than appearing as an element within it.
An alternative would be to let \texttt{FPD\_Process} also inherit from \texttt{FPD\_Object} and carry the same shared attributes.
The separate inheritance path was chosen to make this structural distinction explicit in the type system and to enable \ac{aml} tools to recognize processes as containers.

\subsection{InterfaceClassLib -- Connection Semantics}

\begin{figure}[htbp]
\centering
\begin{subfigure}[t]{0.48\columnwidth}
\centering
\includegraphics[width=\textwidth]{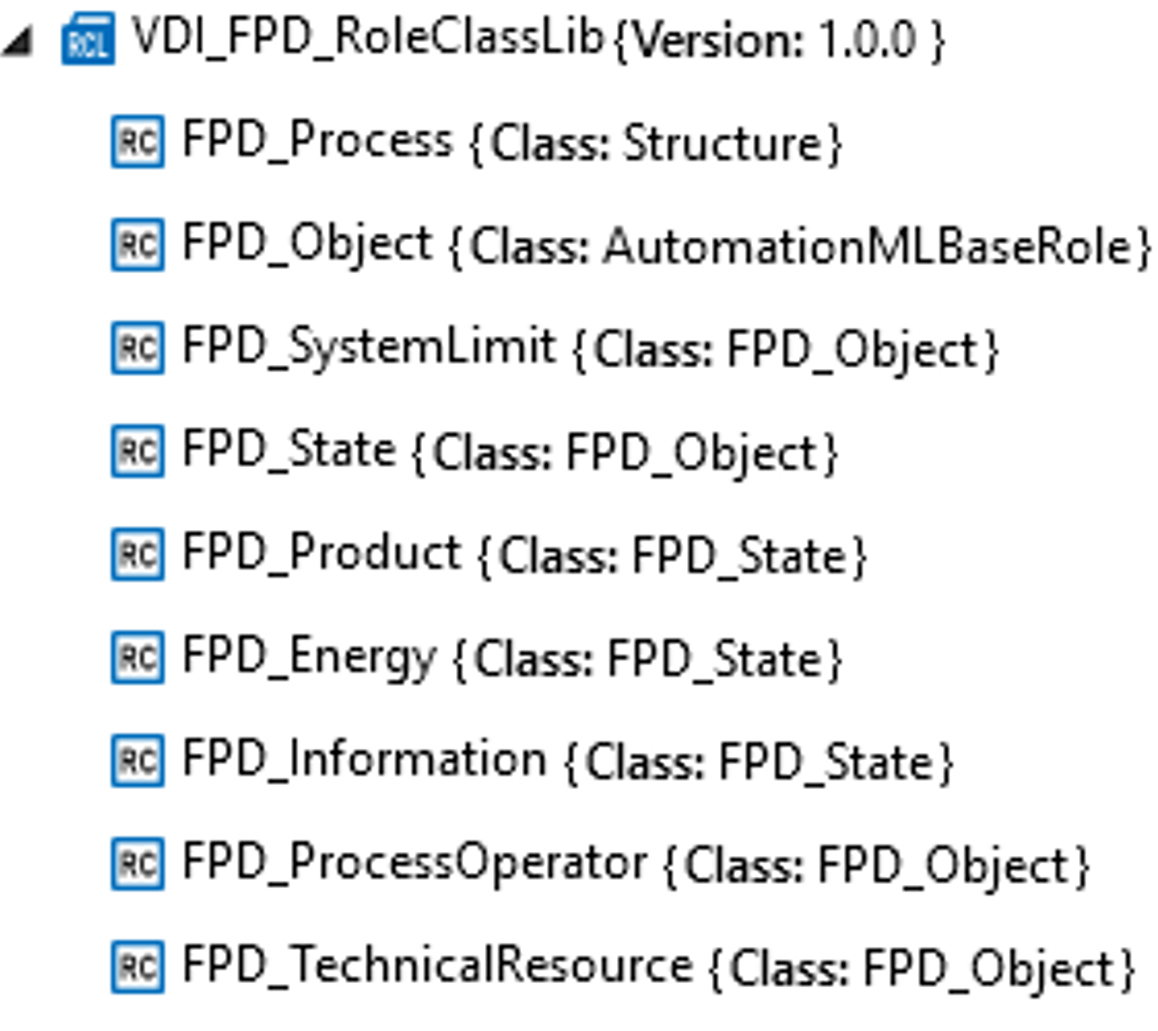}
\caption{VDI\_FPD\_RoleClassLib.}
\label{fig:roleclass}
\end{subfigure}
\hfill
\begin{subfigure}[t]{0.48\columnwidth}
\centering
\includegraphics[width=\textwidth]{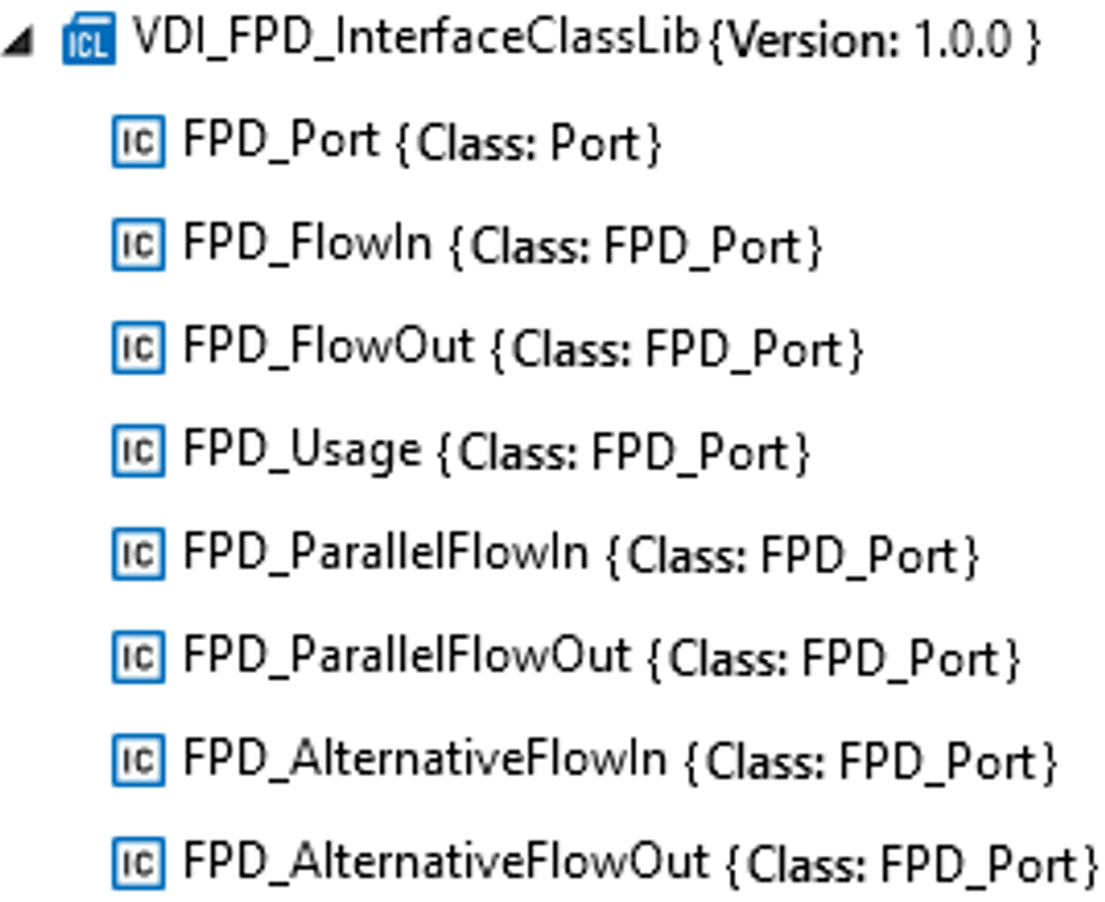}
\caption{VDI\_FPD\_InterfaceClassLib.}
\label{fig:icl}
\end{subfigure}
\caption{Role and interface class libraries. The \texttt{Class} annotation in (a) indicates the parent class from which each role inherits; all interface classes in (b) inherit from \texttt{FPD\_Port}, which itself inherits from the \acs{aml} base interface \texttt{Port}.}
\label{fig:rcl_icl}
\end{figure}

The \textit{VDI\_FPD\_InterfaceClassLib} (Fig.~\ref{fig:icl}) models directed connections as interface pairs.
\texttt{FPD\_Port} inherits from the \ac{aml} base interface \texttt{AutomationMLBaseInterface/Port} and serves as the base class. 
It carries a \texttt{PortCoordinate} attribute (of type \texttt{FPD\_Point}) that stores the position of the connection point on the element.
Since \ac{caex} InternalLinks do not carry direction semantics, flow direction is encoded through typed interface pairs: \texttt{FPD\_FlowIn}/\texttt{Out}, \texttt{FPD\_ParallelFlowIn}/\texttt{Out}, and \texttt{FPD\_AlternativeFlowIn}/\texttt{Out}.
Usage relationships are modeled via a single \texttt{FPD\_Usage} interface rather than an input/output pair, since the standard defines usage as an undirected association between a process operator and a technical resource.
An alternative would be to use a single interface type and encode direction as an attribute; the typed-pair approach was chosen because it allows \ac{aml} tools to determine flow direction purely from the interface types, without inspecting attribute values.
Connections are realized as \texttt{InternalLinks} between ExternalInterface instances on the connected elements.

\subsection{AttributeTypeLibs}

\begin{figure}[htbp]
\centering
\begin{subfigure}[t]{0.48\columnwidth}
\centering
\includegraphics[width=\textwidth]{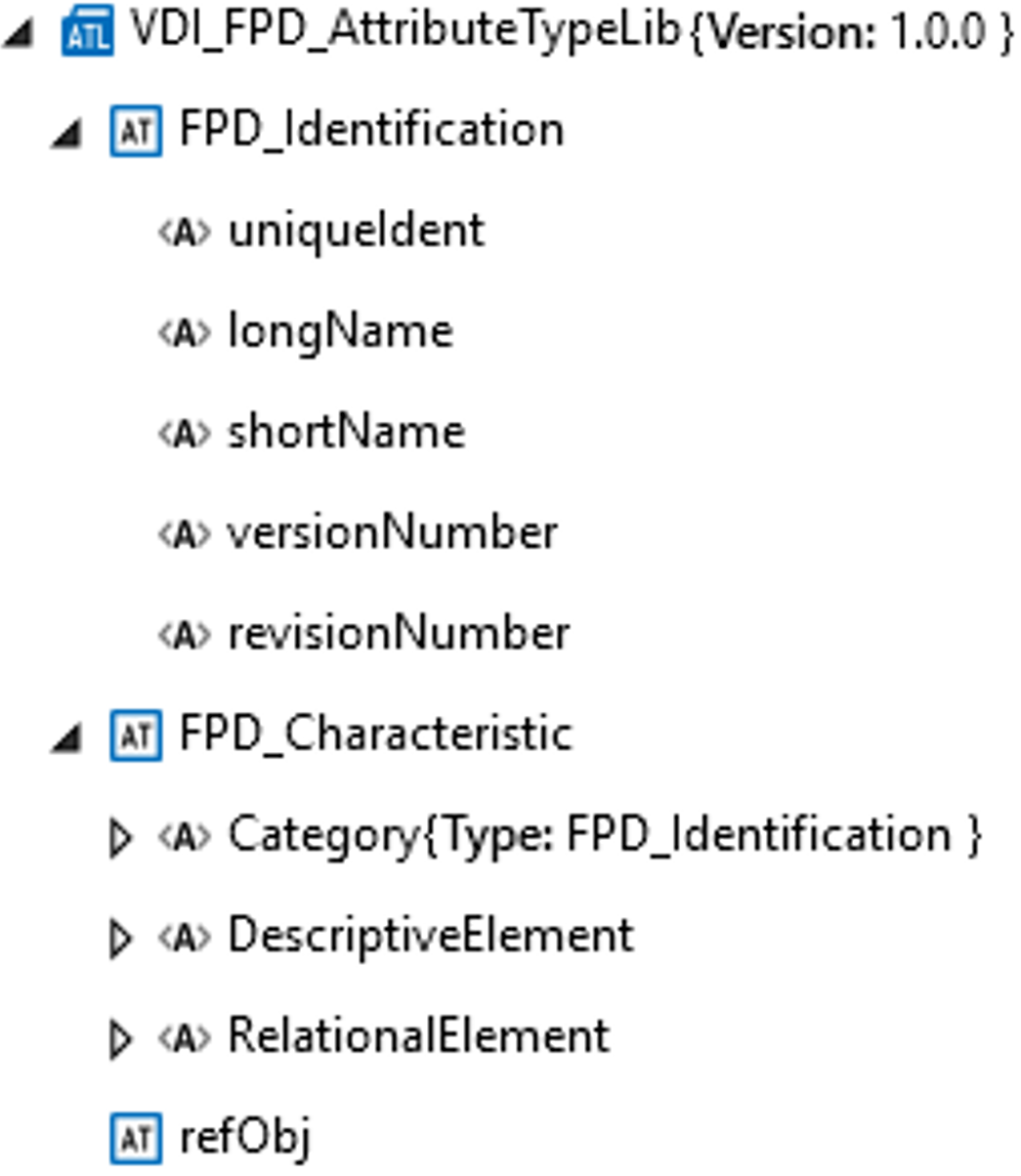}
\caption{VDI\_FPD\_AttributeTypeLib.}
\label{fig:atl}
\end{subfigure}
\hfill
\begin{subfigure}[t]{0.48\columnwidth}
\centering
\includegraphics[width=\textwidth]{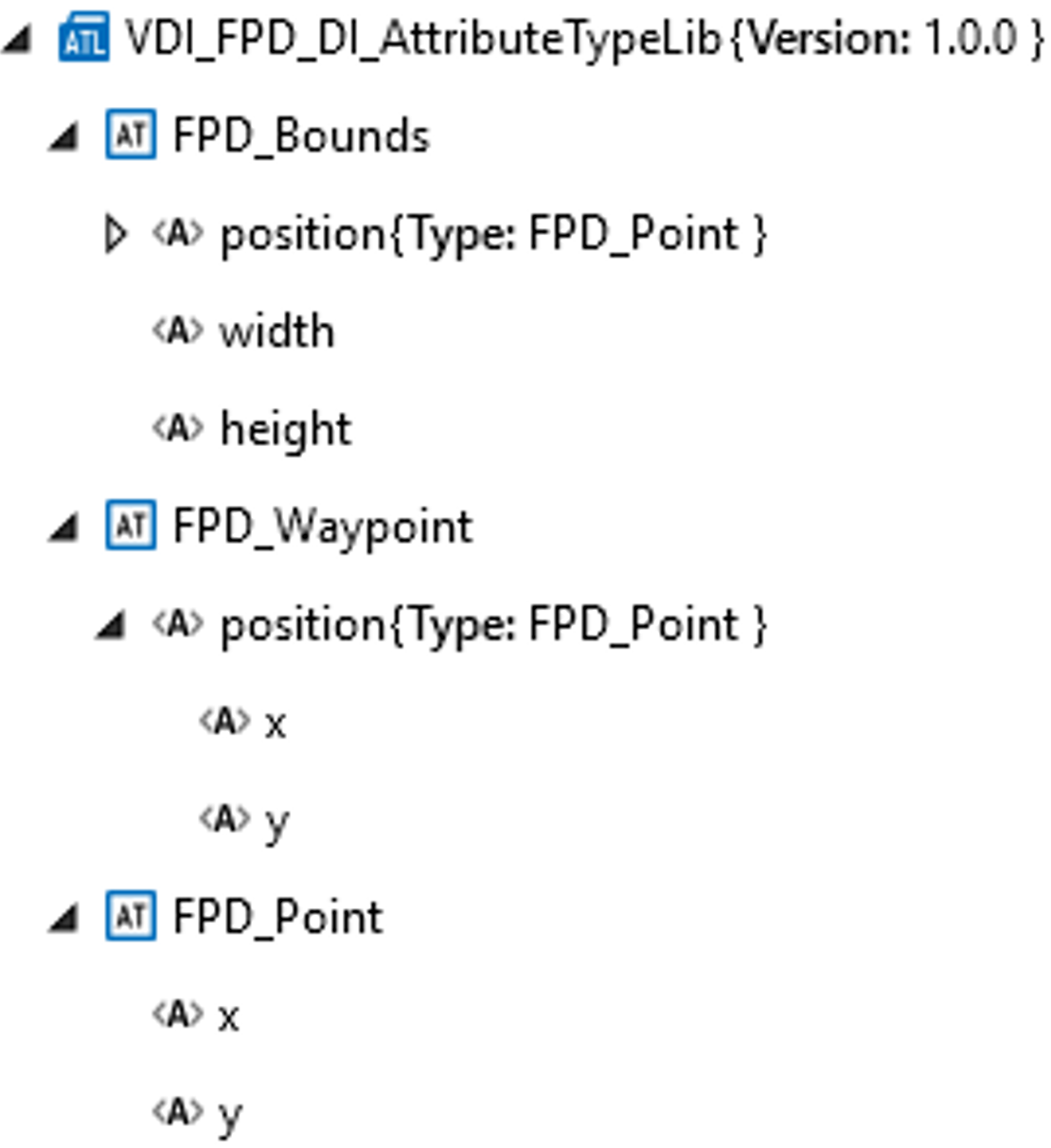}
\caption{VDI\_FPD\_DI\_AttributeTypeLib.}
\label{fig:diatl}
\end{subfigure}
\caption{Attribute type libraries. (a) defines \texttt{FPD\_Identification}, \texttt{FPD\_Characteristic}, and the \texttt{refObj} reference attribute; (b) defines graphical representation attributes based on the OMG \acs{dd} metamodel.}
\label{fig:atls}
\end{figure}

Two AttributeTypeLibs formalize the data structures.
The \textbf{VDI\_FPD\_AttributeTypeLib} (Fig.~\ref{fig:atl}) defines \texttt{FPD\_Identification} (with \textit{uniqueIdent}, \textit{longName}, \textit{shortName}, \textit{versionNumber}, \textit{revisionNumber}), \texttt{FPD\_Characteristic} (with \textit{Category}, \textit{DescriptiveElement}, \textit{RelationalElement} per~\cite{VDI3682Blatt2}), and a generic \texttt{refObj} reference attribute used for cross-diagram references such as decomposition links and boundary state mappings. The mechanism relates to broader multi-context reference concepts proposed for AutomationML~\cite{Drath2026ETFA}.
The \textbf{VDI\_FPD\_DI\_AttributeTypeLib} (Fig.~\ref{fig:diatl}) defines graphical representation attributes based on the OMG \ac{dd} metamodel~\cite{OMGDD}, which was chosen because its data types are already established in standards such as BPMN and UML. The library defines \texttt{FPD\_Bounds} (position, width, height), \texttt{FPD\_Waypoint} (routing point), and \texttt{FPD\_Point} (x/y coordinate).
Since the \ac{fpd} does not treat connections as first-class object types with their own identity~\cite{VDI3682Blatt1}, waypoints are stored on the outgoing ExternalInterface of the element from which the link originates; for usage relationships, this is conventionally the process operator.
An alternative would be to model connections as separate InternalElements with their own graphical attributes, which would require extending the information model beyond the current standard.

\subsection{SystemUnitClassLib and Instantiation}

\begin{figure}[htbp]
\centering
\includegraphics[width=0.82\columnwidth]{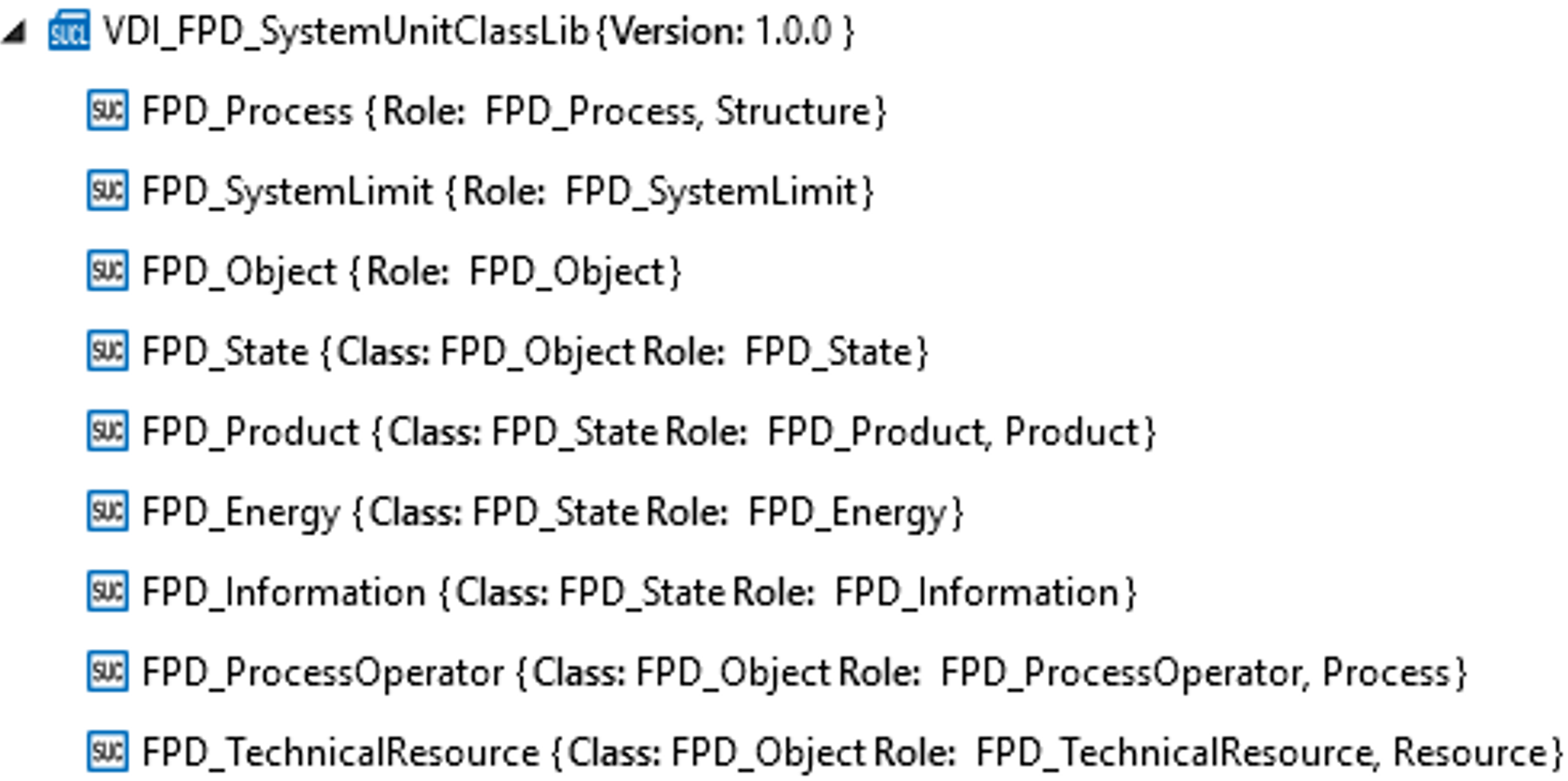}
\caption{VDI\_FPD\_SystemUnitClassLib. Each entry shows the \acs{caex} class inheritance (\texttt{Class}) and role mappings (\texttt{Role}).}
\label{fig:suc}
\end{figure}

The \textit{VDI\_FPD\_SystemUnitClassLib} (Fig.~\ref{fig:suc}) provides instantiation templates corresponding to the RoleClasses.
Each class references its RoleClass via \texttt{SupportedRoleClass}.
Where a matching concept exists in the \ac{aml} base role library, classes additionally declare a second role mapping:
\begin{itemize}
\item \texttt{FPD\_Product} supports \texttt{AutomationMLBaseRole/Product},
\item \texttt{FPD\_ProcessOperator} supports \texttt{AutomationMLBaseRole/Process},
\item \texttt{FPD\_TechnicalResource} supports \texttt{AutomationMLBaseRole/Resource}, and
\item \texttt{FPD\_Process} supports \texttt{AutomationMLBaseRole/Structure}.
\end{itemize}
Classes such as \texttt{FPD\_Energy}, \texttt{FPD\_Information}, and \texttt{FPD\_SystemLimit} have no generic \ac{aml} counterpart and are therefore mapped only to their domain-specific roles.
This dual-role mapping enables \ac{aml} tools to interpret \ac{fpd} elements both in their domain-specific semantics and in terms of generic \ac{aml} concepts---for example, an \ac{aml} tool that queries all elements with the role \texttt{Resource} will find \ac{fpd} technical resources without requiring \ac{fpd}-specific knowledge.

Two key design decisions govern how concrete models are instantiated. 
A concrete \ac{fpd} model is an \texttt{InstanceHierarchy} containing \texttt{FPD\_Process} instances, each representing a diagram with its own coordinate space.
Within a process, all object types---system limit, states, process operators, and technical resources---are modeled as direct children at the same hierarchy level. 
The system limit does not serve as a \ac{caex} container for the other elements, regardless of their graphical placement inside or outside it.
An alternative would be to nest elements inside the system limit, reflecting the graphical containment.
The flat structure was chosen because the system limit in the \ac{fpd} is a boundary marker rather than a semantic container; nesting would imply a containment relationship that the standard does not define.

Hierarchical decomposition is realized through reference attributes rather than \ac{caex} element nesting.
On a process operator, a \texttt{refProcess} attribute (a named specialization of the generic \texttt{refObj} attribute type) points to the child process; on a child process, \texttt{refObj} points back to the parent operator; on a boundary state, \texttt{refObj} references the corresponding top-level state.
An alternative would be to nest child processes inside their parent process operators.
The reference-based approach was chosen because it preserves independent coordinate spaces per diagram and supports arbitrary decomposition depth without deeply nested \ac{caex} structures.

\section{Validation}
\label{sec:validation}
\begin{figure*}[htbp]
\centering
\includegraphics[width=0.82\textwidth]{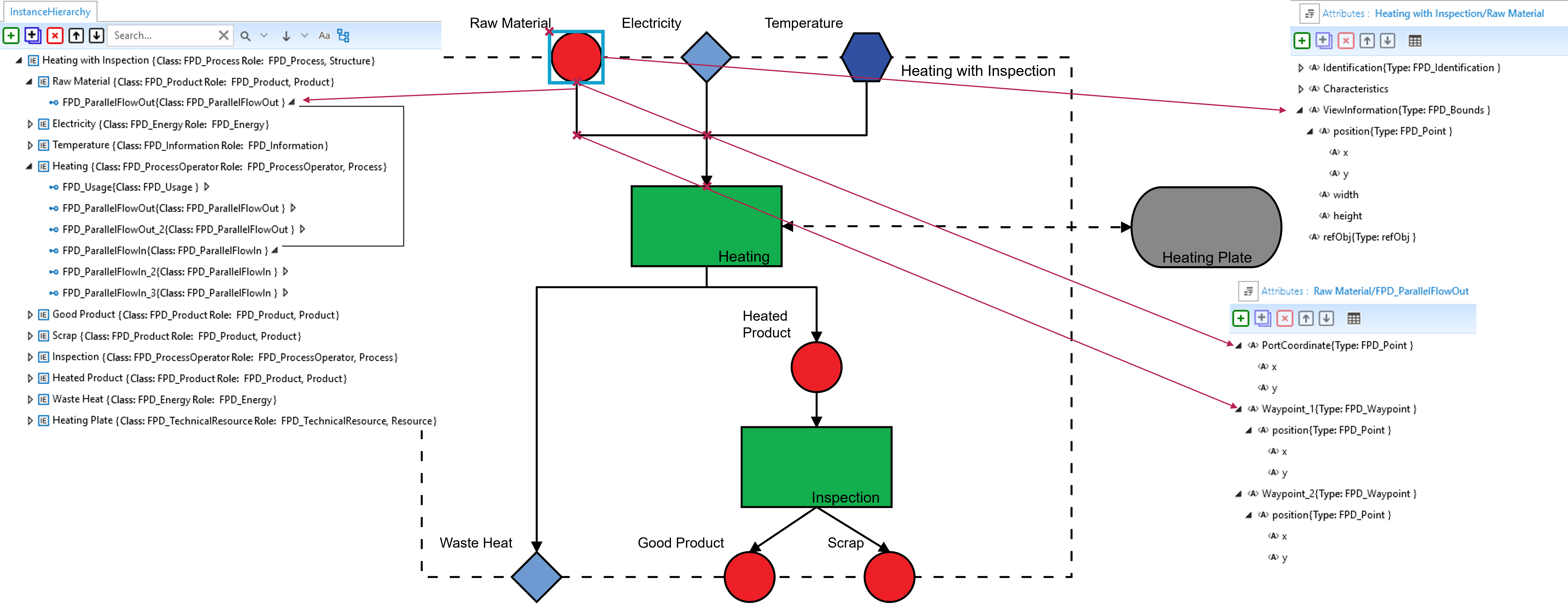}
\caption{The process ``Heating with Inspection'' modeled in FPB.js (\fpbjsurl, center) and exchanged via the domain library to the \acs{aml} Editor, showing instance hierarchy (left) and element attributes (right).}
\label{fig:example}
\end{figure*}

\urldef{\fpbjsurl}\url{https://www.fpbjs.net}
The library was validated using a bidirectional mapping tool\footnote{Available at \url{https://aml.fpbjs.net}\label{fn:amlurl}} built on the \ac{aml} Engine SDK (.NET) that converts \ac{fpd} models between the web-based modeler FPB.js~\cite{Nabizada2020} and the \ac{aml} domain library.
Validation was conducted on multiple models of varying complexity, including models with all six object types, all three flow types plus usage relationships, and multi-level hierarchical decomposition with boundary states.
Fig.~\ref{fig:example} shows a representative example: the process ``Heating with Inspection'' as displayed in the \ac{aml} Editor, including the \ac{caex} instance hierarchy, the graphical diagram, and the attributes of a selected element. The complete example including all decomposition layers is available online.\footref{fn:amlurl}

For the evaluated models, the mapping was verified to be lossless in both directions: all objects, attributes (Identification, Characteristics), connection topologies, decomposition links, and graphical representation (element positions, dimensions, connection waypoints) were preserved through a round-trip conversion (FPB.js $\rightarrow$ AML $\rightarrow$ FPB.js). 
Verification was performed by an automated comparison of the source and re-imported models.
The resulting \ac{aml} files were additionally opened in the \ac{aml} Editor to verify that the object hierarchy, role assignments, and attribute values are correctly represented and navigable without requiring \ac{fpd}-specific editor plugins or rendering logic.
A limitation of the current validation is that round-trip conversion was only demonstrated with a single tool (FPB.js). Validation with additional, independently developed \ac{fpd} tools would further strengthen the interoperability claim.

\section{Conclusion}

This paper presented an \ac{aml} domain library that formalizes the \ac{fpd} according to VDI/VDE~3682 as \ac{caex}~3.0 class libraries.
The library covers all object types, their attributes, four connection types, hierarchical decomposition, and graphical representation based on the OMG \ac{dd} metamodel.
Key design decisions were discussed along with the alternatives that were considered.
Since the library builds on \ac{aml}, \ac{fpd} models can be embedded alongside other engineering artifacts---such as plant structure models or control configurations---within a single \ac{aml} project, enabling cross-domain reuse without additional integration effort.

As the library is a candidate for Part~3 of VDI/VDE~3682, the design decisions are documented here to support the ongoing standardization in GMA Technical Committee~2.19 and to enable community feedback. 
The proposed library, an example, and a feedback function are available online.\footref{fn:amlurl}
Future work includes the finalization of the library within the standardization process, validation with additional independently developed tools, the application of the library in industrial engineering workflows, and the alignment of the \texttt{refObj} mechanism with the generalized multi-context reference concept proposed for AutomationML~\cite{Drath2026ETFA}. 
In addition, an \ac{aml} Editor plugin for native \ac{fpd} modelling is under development.\footnote{\url{https://github.com/hsu-aut/AMLFPB.js}}

\section*{Acknowledgment}

Parts of this work were funded by dtec.bw [project iMOD] -- Digitalization and Technology Research Center of the Bundeswehr. dtec.bw is funded by the European Union -- NextGenerationEU.

{\footnotesize
\bibliographystyle{IEEEtran}
\bibliography{references}
}

\end{document}